\documentclass[aps,prl,twocolumn,amsmath,amssymb,showpacs,letterpaper]{revtex4}

\usepackage{graphicx}
\usepackage{bm}
\usepackage{amsmath}
\usepackage{mathtools}
\usepackage{SIunits}
\usepackage{psfrag}
\usepackage{pstricks}
\usepackage{xspace}
\usepackage{color}
\usepackage[normalem]{ulem}

\newcommand{\ie}{i.e.\@\xspace}

\newcommand{\fig}[1]{Fig.~\ref{#1}}

\renewcommand{\bm}[1]{\boldsymbol{\mathbf{#1}}}
\newcommand{\ud}{\mathrm{d}}
\newcommand{\bra}{\left\langle}
\newcommand{\ket}{\right\rangle}

\newcommand{\hankel}{\operatorname{H}_0^{(1)}}
\newcommand{\bessel}{\operatorname{J}_0}

\newcommand{\NoAutoSpaceBeforeFDP}[0]{}
\newcommand{\AutoSpaceBeforeFDP}[0]{}
\newcommand{\noautospacebeforefdp}[0]{}
\newcommand{\autospacebeforefdp}[0]{}

\newcounter{tempa}
\newcounter{tempb}
\newcounter{tempc}
\newcounter{tempd}

\newenvironment{diagc}[1]{\psset{unit=1.5mm,fillstyle=solid,fillcolor=white}
   \begin{pspicture}[shift=0.0](0,-1)(#1,2)}{\end{pspicture}
}

\newcommand{\correldeux}[2]{
   \setcounter{tempa}{#2}
   \addtocounter{tempa}{-#1}
   \divide \value{tempa} by 2
   \setcounter{tempb}{#1}
   \addtocounter{tempb}{#2}
   \divide \value{tempb} by 2
   \psarc[linestyle=dashed](\value{tempb},0){\value{tempa}}{0}{180}
}

\newenvironment{ddiag}[1]{\psset{unit=1.5mm,fillstyle=solid,fillcolor=white}
   \begin{pspicture}[shift=-5](0,-6)(#1,6)}{\end{pspicture}
}

\newcommand{\pparticule}[2]{\pscircle(#1,#2){1}}

\newcommand{\iidentique}[4]{
   \psline(#1,#2)(#3,#4)
}
\newcommand{\ccorreldeuxa}[2]{
   \setcounter{tempa}{#2}
   \addtocounter{tempa}{-#1}
   \divide \value{tempa} by 2
   \setcounter{tempb}{#1}
   \addtocounter{tempb}{#2}
   \divide \value{tempb} by 2
   \psarc[linestyle=dashed](\value{tempb},3){\value{tempa}}{0}{180}
}

\newcommand{\ccorreldeuxb}[2]{
   \setcounter{tempa}{#2}
   \addtocounter{tempa}{-#1}
   \divide \value{tempa} by 2
   \setcounter{tempb}{#1}
   \addtocounter{tempb}{#2}
   \divide \value{tempb} by 2
   \psarc[linestyle=dashed](\value{tempb},-3){\value{tempa}}{180}{360}
}
\newcommand{\ccorreldeuxc}[4]{
   \psline[linestyle=dashed](#1,#2)(#3,#4)
}

\begin{document}

\title{High-density hyperuniform materials can be transparent}

\author{O. Leseur}
\affiliation{ESPCI ParisTech, PSL Research University, CNRS, Institut Langevin, 1 rue Jussieu, F-75005, Paris, France}
\author{R. Pierrat}
\affiliation{ESPCI ParisTech, PSL Research University, CNRS, Institut Langevin, 1 rue Jussieu, F-75005, Paris, France}
\author{R. Carminati}\email{remi.carminati@espci.fr}
\affiliation{ESPCI ParisTech, PSL Research University, CNRS, Institut Langevin, 1 rue Jussieu, F-75005, Paris, France}
\date{\today}

\begin{abstract}
We show that materials made of scatterers distributed on a stealth hyperuniform point pattern can be transparent at densities
for which an uncorrelated disordered material would be opaque due to multiple scattering. The conditions for transparency
are analyzed using numerical simulations, and an explicit criterion is found based on a perturbative theory.
The broad applicability of the concept offers perspectives for various applications in photonics, and more generally in
wave physics.
\end{abstract}


\maketitle

% Introduction
% ============
\section{Introduction}

% Main introduction
The study of light propagation in scattering media has been a very active field in the past decades, stimulated by
fundamental questions in mesoscopic physics~\cite{SHENG-1995,MONTAMBAUX-2007} and by the development of innovative
imaging techniques~\cite{SEBBAH-2001}. Recently, a new trend has emerged, with the possibility to control
electromagnetic wave propagation in disordered media up to the optical frequency range.  On the one hand, wavefront
shaping techniques offer the possibility to overcome the distorsions induced by a scattering material, even in the
multiple scattering regime~\cite{VELLEKOOP-2007,CARMINATI-2010-4,FINK-2012}. On the other hand, the possibility to
engineer the disorder itself, by controlling the degree of structural correlation, opens new perspectives for the design
of materials with specific properties (e.g., absorbers or filters for
photonics)~\cite{SCHEFFOLD-2004,REUFER-2007,CORTE-2008,LOPEZ-2007,CEFE-LOPEZ-2010}. These materials combine the advantages of
disordered materials, in terms of process scalability and robustness to fabrication errors, with the possibility to
develop a real engineering of their scattering and transport properties through the control of the degree of correlation
in the disorder. For example, it has been shown that correlations can substantially change basic transport properties,
such as the mean-free path~\cite{MARET-1990},
the density of states~\cite{CAO-2010-4,CAO-2003-1} including
the appearance of bandgaps~\cite{EDAGAWA-2008,SEGEV-2011,ZHANG-2001}, or the Anderson localization
length~\cite{CONLEY-2014}.

% Introduction about correlations and hyperuniformity
A specific class of correlated materials has appeared recently, initially referred to as
``superhomogeneous materials''~\cite{GABRIELLI-2002}, and now called ``hyperuniform materials''~\cite{TORQUATO-2003}.
These materials are made of discrete scatterers distributed on a hyperuniform point pattern, a correlated pattern
with a structure factor $S(\bm{q})$ vanishing in the neighborhood of $|\bm{q}|=0$.
The geometrical properties of hyperuniform point patterns have been extensively studied, in particular in terms of packing
properties~\cite{TORQUATO-2004,DONEV-2005,TORQUATO-2008,TORQUATO-2011,DREYFUS-2015}.
Regarding wave propagation, it has been shown that bandgaps could be observed for electromagnetic waves in two-dimensional (2D)
disordered hyperuniform materials~\cite{TORQUATO-2009,CHAIKIN-2013,MULLER-2013,TORQUATO-2013,TORQUATO-2013-2}. Although
understanding the origin of the bandgaps is still a matter of study~\cite{AMOAH-2015,SCHEFFOLD-2016}, these
results have stimulated the design and fabrication of three-dimensional (3D) hyperuniform structures for wave control at optical
frequencies~\cite{SCHEFFOLD-2013,HABERKO-2013}.

% Introduction to this work
In this Letter, we demonstrate that stealth hyperuniform point patterns, a special class of hyperuniform structures for which
$S(\bm{q})=0$ in a finite domain around $|\bm{q}|=0$, offer the possibility to design disordered materials
that can be both dense and transparent, in a specific and broad range of frequencies and directions of incidence.
The analysis is based on full numerical simulations and theoretical modelling. In the single scattering regime, transparency
can be explained in simple terms, as a direct consequence of the vanishing of the structure factor. Interestingly,
transparency can survive in the multiple scattering regime, under a general condition that we establish using a theoretical analysis
that applies to a broad range of materials.

%Hyperuniform point patterns
% ============
\section{Hyperuniform point patterns}

A distribution of points in space is hyperuniform when the normalized variance of the number of points within a sphere of radius $R$ vanishes
when $R$ tends to infinity~\cite{TORQUATO-2003}. In other words, the fluctuations in the number of points in a volume
increase slower with the volume than the average number of points, a feature at the origin of the name ``hyperuniform''. In Fourier space, this is
equivalent to stating that the structure factor defined as~\cite{ZIMAN-1979,HANSEN-2005}
\begin{equation}\label{structure_factor}
   S(\bm{q})=\frac{1}{N}\left\lvert\sum_{j=1}^N \exp[i\bm{q}\cdot\bm{r}_j]\right\rvert^2
\end{equation}
vanishes in the neighborhood of $|\bm{q}|=0$~\cite{TORQUATO-2008}, with $N$ the number of
points and $\bm{r}_j$ their positions. In this work we study scattering of light or other electromagnetic waves by disordered
materials made of discrete scatterers whose positions coincide with a hyperuniform point pattern. 
In the present work, we restrict the numerical study to 2D materials, which is convenient for the sake of computer time, while the theory 
developed in the last section covers both 2D and 3D geometries. Although the design of 3D
materials is a major goal, the interest of 2D architectures should not be underestimated. Artificial 2D structures can produce efficient
microwave reflectors or filters~\cite{CHAIKIN-2013}, and promising 2D materials for photonics and optoelectronics can be fabricated using bottom-up
approaches~\cite{TIAN-2011}. Even natural
materials, as the cornea, can exhibit photonic properties resulting from an underlying 2D
microstructure~\cite{FARRELL-1976}. 

\begin{figure}
   \psfrag{0}{\tiny $0$}
   \psfrag{1}{\tiny $1$}
   \psfrag{2}{\tiny $2$}
   \psfrag{3}{\tiny $3$}
   \psfrag{k1}[c]{(a)}
   \psfrag{k2}[c]{(c)}
   \psfrag{l1}[c]{(b)}
   \psfrag{l2}[c]{(d)}
   \psfrag{x}[c]{\small $x$}
   \psfrag{y}[t]{\small $y$}
   \psfrag{kx}[t]{\small $q_x$}
   \psfrag{ky}[c]{\small $q_y$}
   \psfrag{O}[c]{\textcolor{white}{$\Omega$}}
   \psfrag{K}[c]{\textcolor{white}{$K$}}
   \includegraphics[width=0.46\linewidth]{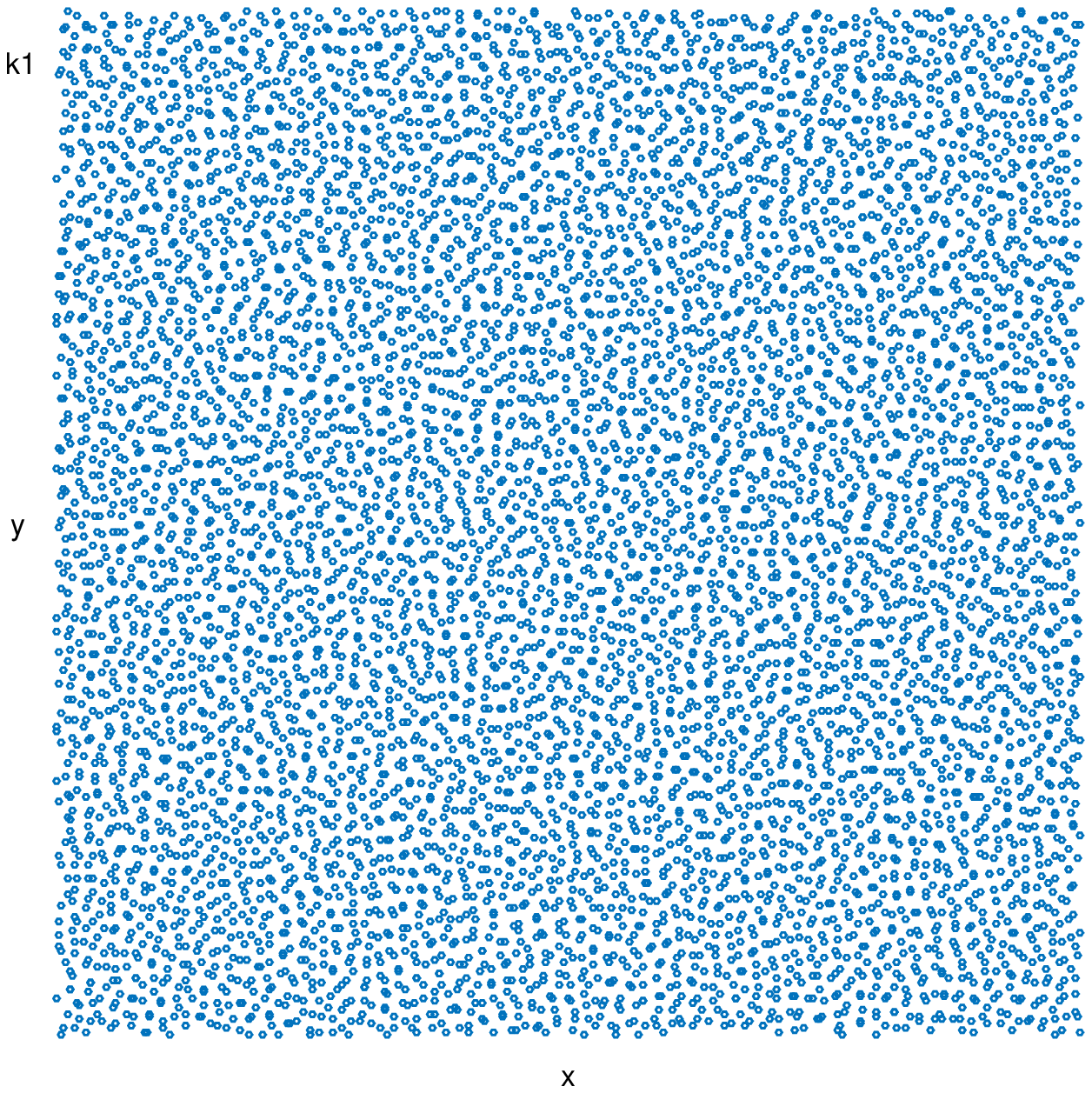}\hfill\includegraphics[width=0.46\linewidth]{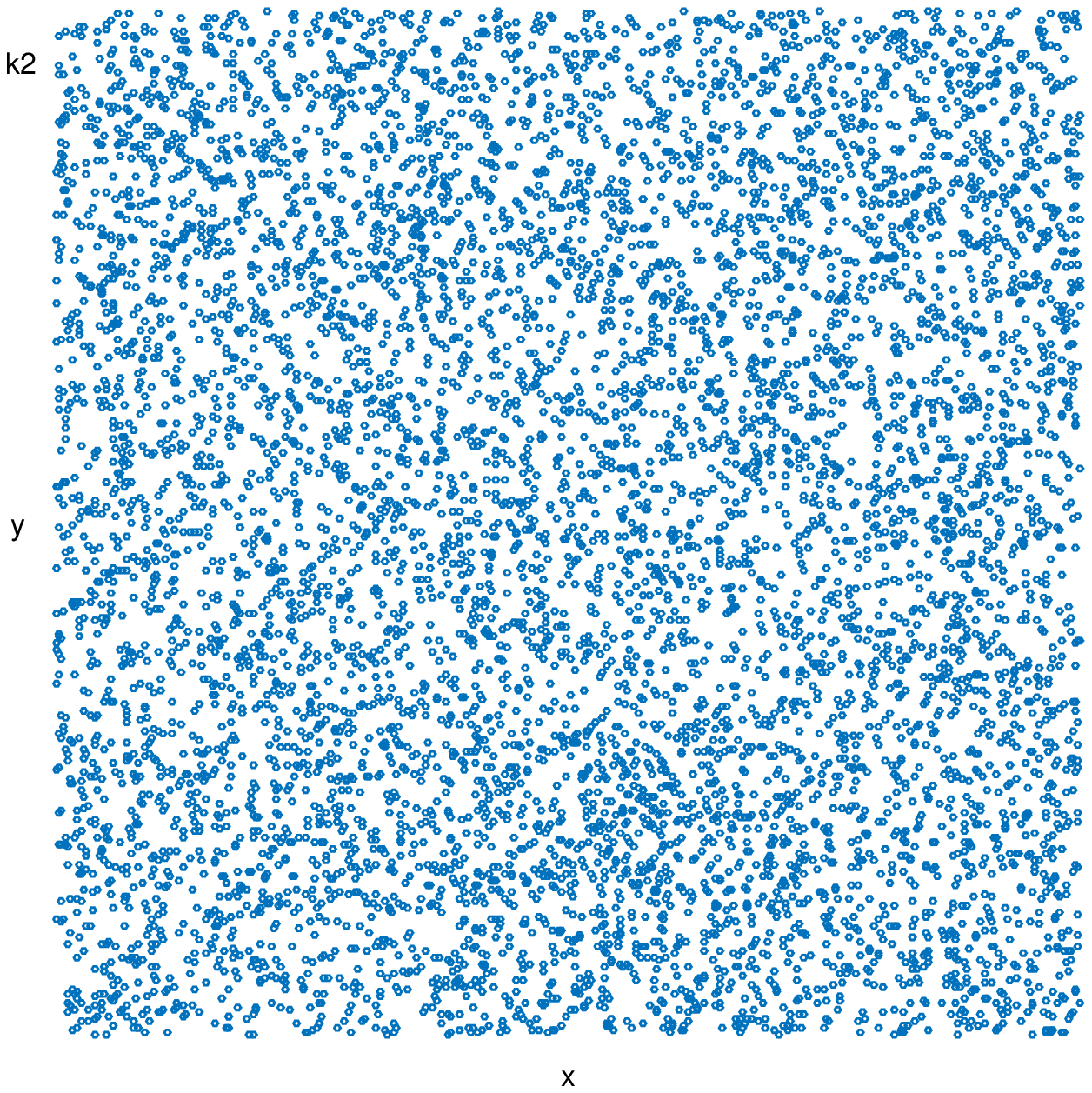}
   \\\includegraphics[width=0.46\linewidth]{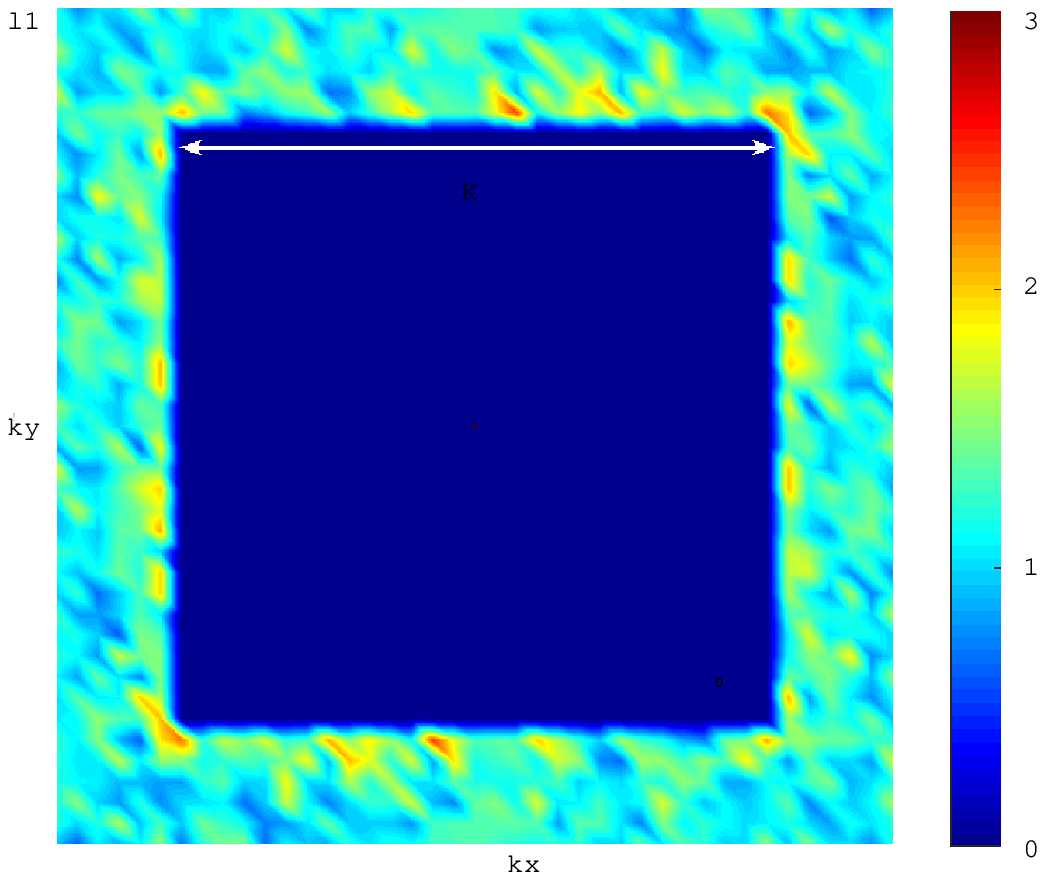}\hfill\includegraphics[width=0.46\linewidth]{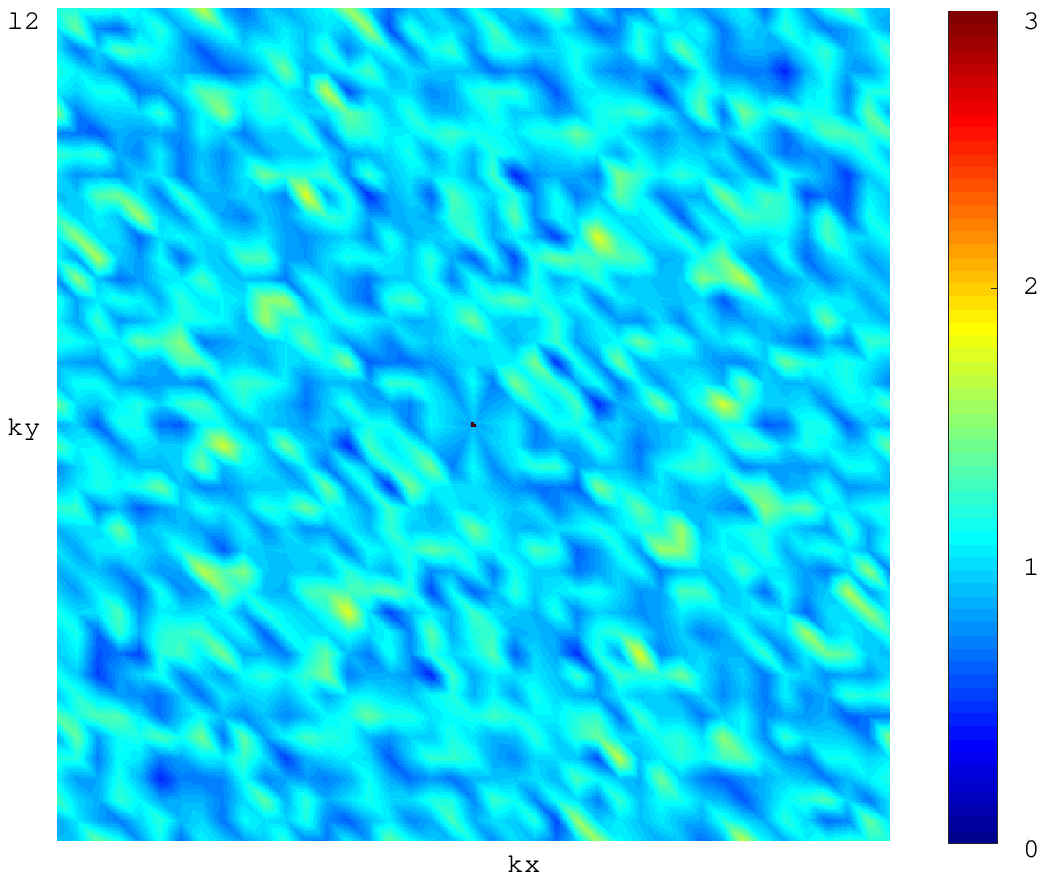}
   \caption{(a) Hyperuniform medium generated with $N=90000$ and $\chi=0.222$.
   (b) Structure factor of the hyperuniform medium averaged over $20$ configurations.
   (c) Uncorrelated medium generated with $N=90000$ and $\chi=0$.
   (d) Structure factor of the uncorrelated medium.}
   \label{structures}
\end{figure}

% Generation of hyperuniform point patterns
The first step consists in generating 2D stealth hyperuniform point patterns. We have adapted the algorithm described in
Refs.~\cite{TORQUATO-2004,TORQUATO-2008-1,UCHE-2006}, and based on
the minimization of an interaction potential (see section~1 of \href{Supplement~1}{link}).
The 2D medium used in the numerical simulations fills a square region with size $L$ and volume $V=L^2$,
with periodic boundary conditions to avoid finite-size effects.
We consider hyperuniform point distributions such that $S(\bm{q})=0$ in a square domain $\Omega$ of reciprocal space with size $K$, 
and centered at the origin. The square shape of $\Omega$ has been chosen for convenience. Although it induces an anisotropy
 in the structure factor, this choice does not affect the generality of the transparency property discussed in this work.
Due to the periodic boundary conditions, the structure factor $S(\bm{q})$ is controlled on a finite number of points located 
on a square lattice inside $\Omega$, and separated by $2\pi/L$.
The number of points $M(K)$ depends on the size $K$ of the domain $\Omega$, and satisfies $2M(K)+1=(K^2L^2)/(4\pi^2)$,
where the factor of $2$ results from the symmetry property $S(\bm{q})=S(-\bm{q})$ and the ``+1'' correction accounts for 
the fact that the value of the structure factor at the exact point $\bm{q}=0$ cannot be controlled [$S(\bm{q}=0)=N$].
For large $K$, the system is very constraint, creating stronger correlations in the point pattern.
The degree of order in hyperuniform structures is usually measured by the parameter $\chi=M(K)/(2N)$.
For $\chi=0$, the system is uncorrelated (fully disordered), while for $\chi=1$, the system is a perfect crystal~\cite{TORQUATO-2004}.
From the expression of $M(K)$, one immediatly gets $K=(2\pi)L^{-1}\sqrt{4N\chi+1}$, showing that by increasing $K$, 
one increases the degree of order in the point pattern.
An example of point distribution generated numerically is shown in \fig{structures}~(a), together with the corresponding
structure factor averaged over 20 configurations in \fig{structures}~(b). In the square area $\Omega$, the
structure factor vanishes [except at the origin where $S(\bm{q}=0)=N$]. For the sake of comparison,
an uncorrelated pattern with the same number of points is shown in \fig{structures}~(c), with the corresponding structure factor
in \fig{structures}~(d). The structure factor of the uncorrelated disordered medium is almost
everywhere close to unity.

% Figure 1: configurations and structure factors

%Scattering
% ============
\section{Scattering from stealth hyperuniform materials}

\begin{figure}
   \psfrag{M}[bl]{}
   \psfrag{t}[cl]{$\theta$}
   \psfrag{L}[t]{$L$}
   \psfrag{K}[c]{$K$}
   \psfrag{O}[c]{$\Omega$}
   \psfrag{w}[c]{$w$}
   \psfrag{q}[c]{$\bm{q}$}
   \psfrag{k}[c]{$\bm{k}_s$}
   \psfrag{l}[c]{$\bm{k}_i$}
   \psfrag{a}[c]{(a)}
   \psfrag{b}[c]{(b)}
   \psfrag{S}[c]{\footnotesize $S(\bm{q})=0$}
   \psfrag{s}[c]{\footnotesize $S(\bm{q})\neq0$}
   \psfrag{ky}[c]{$q_y$}
   \psfrag{kx}[c]{$q_x$}
   \includegraphics[width=0.5\linewidth]{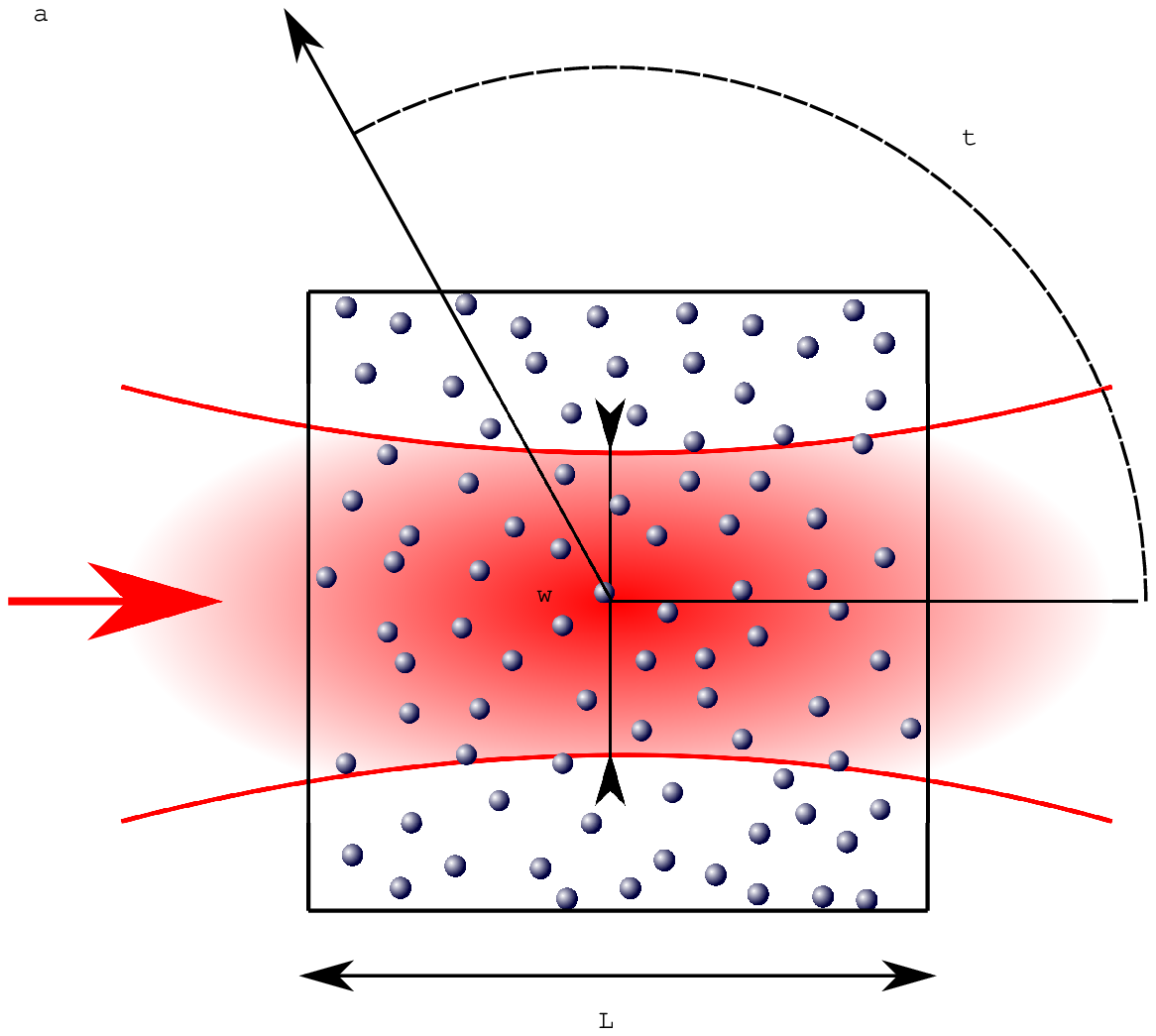}\includegraphics[width=0.5\linewidth]{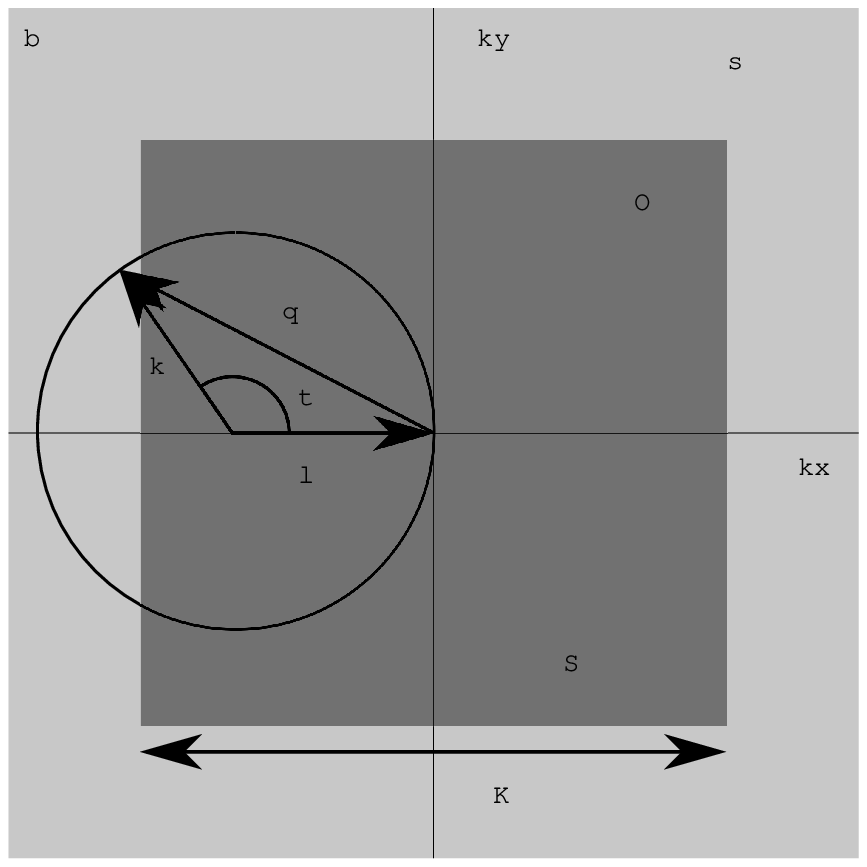}
   \psfrag{d10}{(c)}
   \psfrag{d00}{(d)}
   \psfrag{c10}{}
   \psfrag{c00}{}
   \psfrag{f1}[cl]{\scriptsize $0$}
   \psfrag{f2}{}
   \psfrag{f3}{}
   \psfrag{f4}[cb]{\scriptsize $\pi/2$}
   \psfrag{f5}{}
   \psfrag{f6}{}
   \psfrag{f7}[cc]{\scriptsize $\pi$}
   \psfrag{f8}{}
   \psfrag{f9}{}
   \psfrag{f10}[cc]{\scriptsize $-\pi/2$}
   \psfrag{f11}{}
   \psfrag{f12}{}
   \psfrag{23}[c]{\scriptsize $230$}
   \psfrag{35}[c]{\scriptsize $350$}
   \psfrag{30}[c]{\scriptsize $300$}
   \psfrag{45}[c]{\scriptsize $450$}
   \\\includegraphics[width=0.5\linewidth]{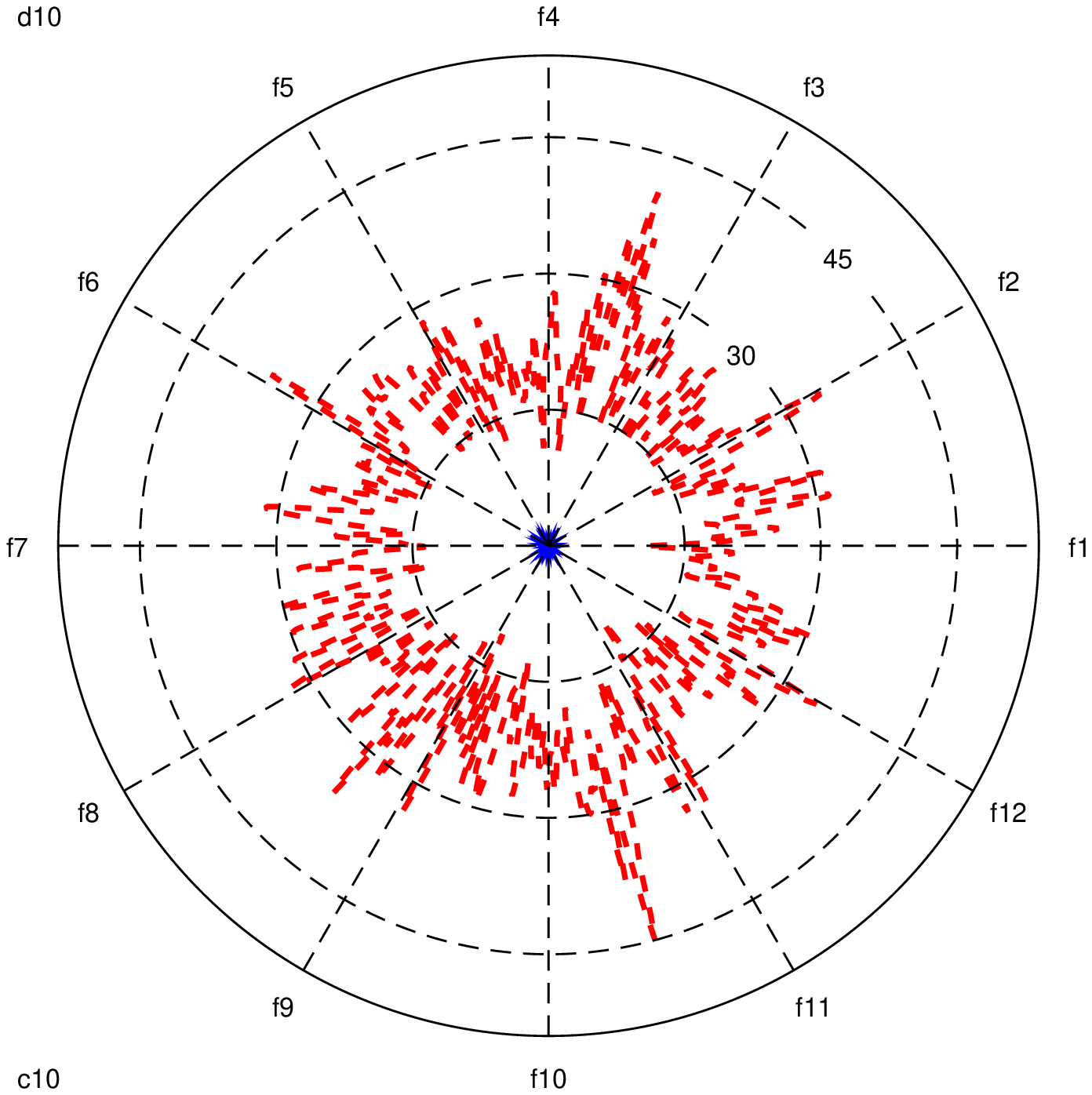}\hfill\includegraphics[width=0.5\linewidth]{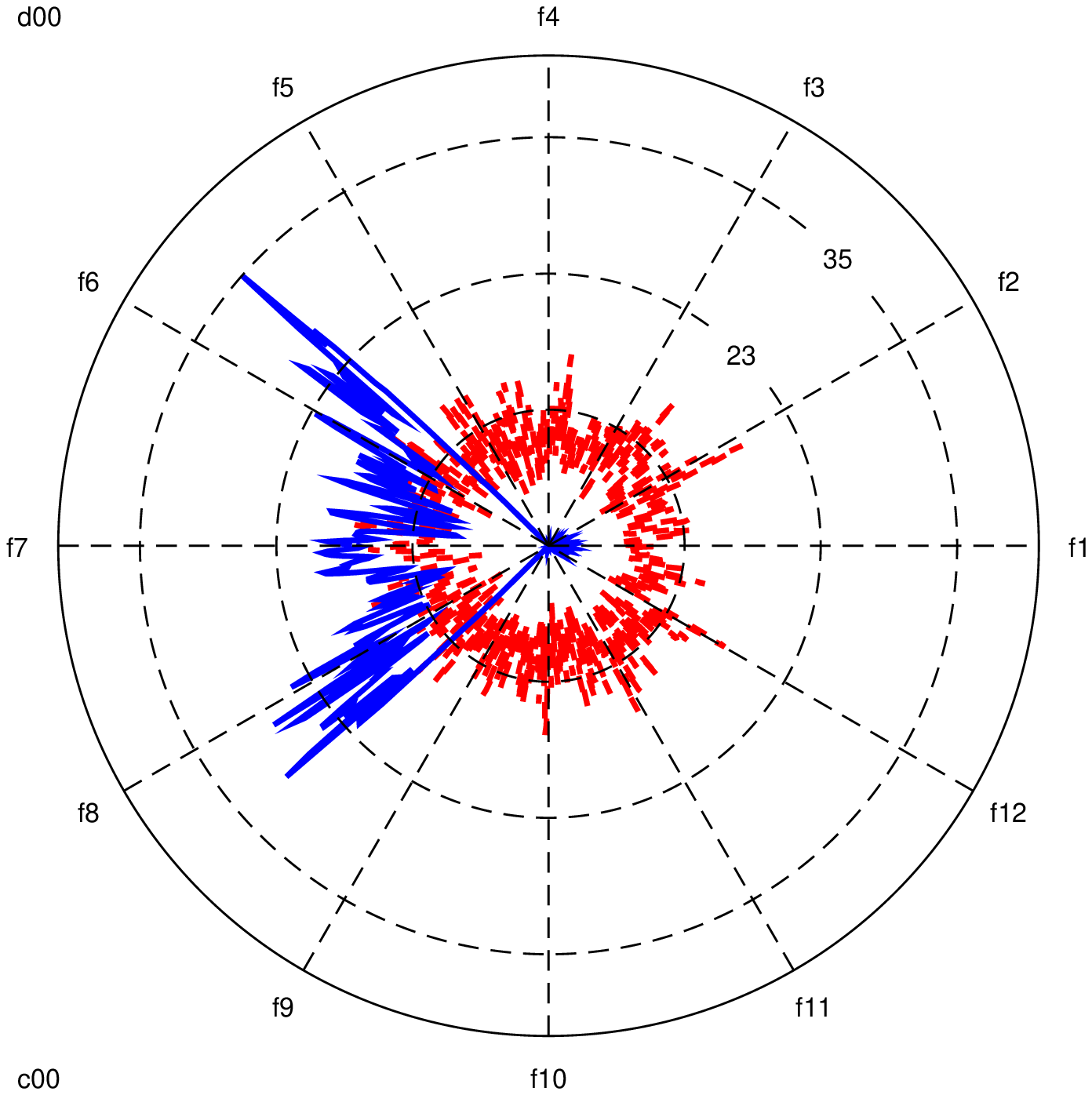}
   \caption{(a) Scattering geometry. The 2D medium is illuminated by a Gaussian beam at normal
   incidence (beam waist $w=60/k_0$), and the diffuse intensity is calculated in direction $\theta$. (b)~Schematic view
   of the domain described by the scattered wavevector $\bm{q}=\bm{k}_s-\bm{k}_i$ in Fourier space. For a hyperuniform
   material, the structure factor vanishes in the square domain $\Omega$.  (c)~Angular pattern of the average diffuse
   intensity in the single-scattering regime ($b_B=0.5$) for a hyperuniform medium with $k_0=K/8$ (blue line) and for
   uncorrelated disorder (red dashed line). The same scatterers and the same density are used in both cases, the
   intensity is averaged over 20 configurations, and $k_0\ell_B=444$. (d) Same as (c) with $k_0=K/3.5$ and
   $k_0\ell_B=1015$.}
   \label{single_scattering}
\end{figure}

% Presentation of the numerical method for light propagation
A hyperuniform scattering material is built by dressing each point with an electric polarizability $\alpha(\omega)$
describing the electrodynamic response of the individual scatterers at frequency $\omega$ (in this study the
scatterers are assumed to be much smaller than the wavelength and treated in the electric-dipole limit).
For 2D waves with an electric field parallel to the invariance axis of the scatterers, the polarizability is
$\alpha(\omega) = -4\gamma c^2/[\omega_0(\omega^2-\omega_0^2+i\gamma\omega^2/\omega_0)]$, where $\omega_0$ is
the resonant frequency and $\gamma$ the linewidth, this expression being consistent with the optical theorem (energy conservation). The
scattering medium is illuminated by a Gaussian beam at normal incidence (direction defined by wavevector $\bm{k}_i$)
and focused in the middle of the medium, as shown in \fig{single_scattering}~(a).
The beam waist $w$ is chosen large compared to the wavelength to mimic a weakly focussed beam similar
to a plane wave. The electric field is calculated numerically by solving Maxwell's equations using the coupled dipoles
method~\cite{LAX-1952}. A description of the method in a similar geometry can be found in Ref.~\cite{LESEUR-2014}. It allows
us to compute the scattered electric field $E_{\text{sca}}(\bm{r})$ at any position inside or outside the medium.
In wave scattering by disordered media, a measurable quantity is the average diffuse intensity, proportionnal to the
average power radiated in a given direction in the far field (defined by wavevector $\bm{k}_s$).
When $k_0 r \gg 1$, $r$ being the observation distance and $k_0=\omega/c=2\pi/\lambda$, with $c$ the speed of light and $\lambda$
the wavelength in vaccum, the scattered field takes the asymptotic form
\begin{equation}
   E_{\text{sca}}(\bm{r},\omega)=\frac{i(1-i)}{4\sqrt{\pi k_0r}}\exp[ik_0r]E_{\text{sca}}(\theta,\omega)
\end{equation}
where $\theta$ is the observation angle (angle between $\bm{k}_i$ and $\bm{k}_s$) indicated in
\fig{single_scattering}~(a).
By definition, the far-field average diffuse intensity is
\begin{equation}\label{average_diffuse_intensity}
   I_{\text{diff}}(\theta,\omega)=\bra \lvert E_{\text{sca}}(\theta,\omega)\rvert^2 \ket - \lvert\bra
         E_{\text{sca}}(\theta,\omega)\ket\rvert^2
\end{equation}
where $\bra\ldots\ket$ denotes an average over the configurations of disorder (positions of scatterers).

% Simulations in the single scattering regime
We first consider the single scattering regime, defined by $b_B(\omega) < 1$, where $b_B(\omega)=L/\ell_B(\omega)$ is the 
optical thickness in the independent scattering (or Boltzmann) approximation, for which the scattering mean free path is 
$\ell_B(\omega)=[\rho\sigma_s(\omega)]^{-1}$ and the condition $k_0\ell_B(\omega) \gg 1$ is satisfied. Here 
$\rho=N/V$ is the number density of scatterers and $\sigma_s(\omega)=(k_0^3/4)|\alpha(\omega)|^2$ their scattering cross section.
Under plane-wave illumination with amplitude $E_0$, it is well-known that the diffuse intensity can be directly written in
terms of the average structure factor. The full expression takes the following form (see section~2 of \href{Supplement~1}{link}):
\begin{equation}\label{intensity_structure_factor}
   I_{\text{diff}}(\theta,\omega)    \propto
         N\bra S(\bm{q})\ket |E_0|^2
         -\frac{N^2}{V^2}\left\lvert\Theta(\bm{q})\right\rvert^2 |E_0|^2
\end{equation}
where $\bm{q}=\bm{k}_s-\bm{k}_i$ is the scattered wavevector. In the right-hand side, $\Theta(\bm{q})$ is the Fourier transform of
a window function equal to unity inside the volume $V$ occupied by the medium and vanishing outside.
This function describes the diffraction pattern of the finite size volume $V$, and in practice is sharply peaked around $\bm{q}=0$.
In particular, for $\bm{q}=0$ corresponding to forward scattering, the two terms in the right-hande side in \eqref{intensity_structure_factor}
exactly compensate each other.

% Figure 2: single scattering regime

In the single scattering regime, the vanishing of $S(\bm{q})$ in a finite domain $\Omega$ suppresses scattering for a given frequency 
and/or angular range, a property at the origin of the denomination stealth hyperuniform materials~\cite{TORQUATO-2008-1}.
To illustrate this interesting behavior, we choose the parameters such
that $b_B=0.5$, and we change the incident wavelength, or equivalently $k_0$, in order to tune the range of scattered wavevector
$\bm{q}=\bm{k}_s-\bm{k}_i=k_0\hat{\bm{k}}_s-k_0\hat{\bm{k}}_i$ that is involved in the scattering process (here $\hat{}$
denotes a unit vector).  The first situation is chosen with $k_0 = K/8$, so that the circle described by the scattered wavevector $\bm{q}$
in \fig{single_scattering}~(b) lies entirely within the domain $\Omega$. In this case we expect a substantial reduction of scattering for
a hyperuniform material [with $S(\bm{q}) \simeq 0$ in $\Omega$], compared to a fully disordered material with the
same density of scatterers. This is observed in \fig{single_scattering}~(c), where we plot the
average diffuse intensity versus the observation angle $\theta$ calculated from the full numerical simulation without approximation
(the averaging is performed over 20 configurations). Also note that this regime is observed as long as the circle 
described by the scattered wavevector $\bm{q}$ lies entirely within the domain $\Omega$, independently on the exact shape of $\Omega$.
For the uncorrelated disordered medium (red dashed line), a diffuse intensity pattern is observed,
while for the hyperuniform structure
made with the same scatterers and the same density, the diffuse intensity (blue line) is negligible [reduced by a factor
of about $800$, so that it is hardly visible in \fig{single_scattering}~(c)].

For smaller wavelengths (larger $k_0$), another behavior can be observed. Choosing $k_0=K/3.5$ as an illustrative example, the circle
described by the scattered wavevector $\bm{q}$ in \fig{single_scattering}~(b) is now only partially included inside the domain $\Omega$.
Therefore, a reduced level of scattering, due to a vanishing structure factor, is expected only for the observation angles
$\theta \in [-\theta_{\text{lim}},+\theta_{\text{lim}}]$ such that $\bm{q}$ falls within $\Omega$. Theoretically, we find
$\theta_{\text{lim}} = \arccos\left[1-K/(2k_0)\right] = \unit{138}{\degree}$. The result of the full numerical simulation of the diffuse
intensity pattern is shown in \fig{single_scattering}~(d). The change between the pattern produced by an uncorrelated disordered
medium (red dashed line) and the hyperuniform material (blue line) is clearly visible. For the latter,
scattering is suppressed for a wide angular range, with a value $\theta_{\text{lim}}=\unit{134}{\degree}$ in good agreement
with the estimate from single-scattering theory (for $b_B=0.5$ higher order scattering is not fully negligible). Finally, let us note
that in terms of frequency bandwidth for a predefined angular range $[-\theta_{\text{lim}},+\theta_{\text{lim}}]$, scattering would be suppressed for all wavelengths
satisfying $\lambda > \lambda_{\text{lim}}$, with $\lambda_{\text{lim}}=(4\pi/K)(1-\cos\theta_{\text{lim}})$.

\begin{figure}
   \psfrag{d11}{(a)}
   \psfrag{dz}{(b)}
   \psfrag{c11}{}
   \psfrag{cz}{}
   \psfrag{f1}[cl]{\scriptsize $0$}
   \psfrag{f2}{}
   \psfrag{f3}{}
   \psfrag{f4}[cb]{\scriptsize $\pi/2$}
   \psfrag{f5}{}
   \psfrag{f6}{}
   \psfrag{f7}[cc]{\scriptsize $\pi$}
   \psfrag{f8}{}
   \psfrag{f9}{}
   \psfrag{f10}[cc]{\scriptsize $-\pi/2$}
   \psfrag{f11}{}
   \psfrag{f12}{}
   \psfrag{10}[c]{\scriptsize $100$}
   \psfrag{15}[c]{\scriptsize $150$}
   \psfrag{28}[c]{\scriptsize $2.8$}
   \psfrag{56}[c]{\scriptsize $5.6$}
   \psfrag{84}[c]{\scriptsize $8.4$}
   \includegraphics[width=\linewidth]{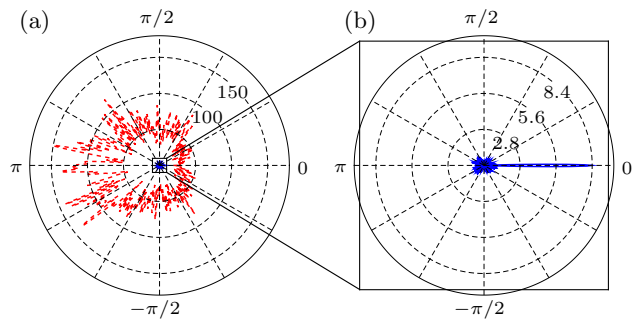}
   \caption{(a) Angular pattern of the average diffuse intensity in the multiple-scattering regime ($b_B=5$) for
   a hyperuniform medium with $k_0=K/8$ (blue line) and for uncorrelated disorder (red dashed line).
   The same scatterers and the same density are used in both cases, the intensity is averaged over 20 configurations,
   and $k_0\ell_B=44.4$. (b) Zoom on the central part of the polar plot.}
   \label{multiple_scattering}
\end{figure}

% Simulations in the multiple scattering regime
The examples above have illustrated the interest of hyperuniform point patterns in the design of materials with a tunable
level of scattering. An interesting question is to see under which conditions these properties could survive in the multiple scattering regime.
The transport of the average intensity in this case is well described by the radiative
transfer equation (RTE)~\cite{CHANDRASEKHAR-1950}, that is derived from the Bethe-Salpeter equation
in the limit $k_0\ell_B\gg1$~\cite{MONTAMBAUX-2007}.
In this framework, the angular dependence of the diffuse intensity is driven by the phase function $p(\bm{u},\bm{u}')$,
where $\bm{u}'$ and $\bm{u}$ are unit vectors denoting the incoming and outgoing directions. For subwavelength scatterers,
the phase function is connected to the average structure factor through the relation (see section~3 of \href{Supplement~1}{link}):
\begin{equation}\label{phase_function}
   p(\bm{u},\bm{u}')\propto\bra S[k_0(\bm{u}-\bm{u}')]\ket
      -\frac{N-1}{V^2}\left\lvert\Theta[k_0(\bm{u}-\bm{u}')]\right\rvert^2
\end{equation}
where the term involving $\Theta$ appears as in  \eqref{intensity_structure_factor}, but with a different weighting factor.
This difference results from the connection between the phase function and the vertex
of the Bethe-Salpeter equation that contains irreducible terms only.
Note that for an uncorrelated disordered medium, the phase function is constant [\ie $p(\bm{u},\bm{u}')=1$].
In the case of a hyperuniform structure, under the condition $k_0<K/4$, the scattered wavevector $k_0(\bm{u}-\bm{u}')$
falls inside the domain $\Omega$ [see the construction in \fig{single_scattering}~(b)], and the
phase function vanishes for all directions, except in the forward direction $\bm{u}=\bm{u}'$. Indeed, the second term
involving $\Theta$ in \eqref{phase_function} does not exactly compensates the value of the structure factor at the origin.
This means that scattering occurs, but only in the forward direction. For an infinite medium, this would lead to an effective
scattering mean free path $\ell$ tending to infinity. In practice, for a medium with size $L$, the width of $\Theta(\bm{k})$
is on the order of $2\pi/L$, keeping $\ell$ finite but permitting in principle to reach arbitary large values.
This is particularly intersting since getting $\ell \gg L$ leads to transparency. To check this idea,
we have used numerical simulations in the multiple scattering regime ($b_B=5$), with $k_0=K/8$ and $k_0\ell_B=44.4$.
The result is displayed in \fig{multiple_scattering}, were we clearly observe that the average diffuse intensity
for the hyperuniform medium (blue line) is extremely small compared to that obtained for a fully disordered medium
(red dashed line), except in the forward direction, as expected. Our analysis demonstrates that, in conditions (in terms of type and
density of scatterers) under which a fully disordered material would induce strong multiple scattering,
a hyperuniform material can chiefly generate forward scattering, leading to an effective
scattering mean free path $\ell \gg L$, which is the condition for transparency.

% Figure 3: multiple scattering regime

%Criterion for transparency
% ============
\section{An explicit criterion for transparency}

It is possible to establish a criterion for the existence of transparency in a dense (stealth) hyperuniform disordered material,
based on a perturbative theory valid for both 2D and 3D materials.
To proceed, we use a diagrammatic expansion of the phase function in the form~\cite{RYTOV-1989}
\begin{equation}\label{phase_function_diagrams}
   p=
   \begin{ddiag}{2}
      \iidentique{1}{3}{1}{-3}
      \pparticule{1}{-3}
      \pparticule{1}{3}
   \end{ddiag}
   +
   \begin{ddiag}{2}
      \ccorreldeuxc{1}{3}{1}{-3}
      \pparticule{1}{-3}
      \pparticule{1}{3}
   \end{ddiag}
   +
   \begin{ddiag}{8}
      \ccorreldeuxa{1}{7}
      \iidentique{1}{3}{1}{-3}
      \iidentique{1}{3}{7}{3}
      \pparticule{1}{-3}
      \pparticule{1}{3}
      \pparticule{7}{3}
   \end{ddiag}
   +      
   \begin{ddiag}{8}
      \ccorreldeuxa{1}{7}
      \iidentique{7}{3}{7}{-3}
      \iidentique{1}{3}{7}{3}
      \pparticule{7}{-3}
      \pparticule{1}{3}
      \pparticule{7}{3}
   \end{ddiag}
   +
   \begin{ddiag}{8}
      \ccorreldeuxb{1}{7}
      \iidentique{1}{3}{1}{-3}
      \iidentique{1}{-3}{7}{-3}
      \pparticule{1}{-3}
      \pparticule{1}{3}
      \pparticule{7}{-3}
   \end{ddiag}      
   + 
   \begin{ddiag}{8}
      \ccorreldeuxb{1}{7}
      \iidentique{7}{3}{7}{-3}
      \iidentique{1}{-3}{7}{-3}
      \pparticule{7}{-3}
      \pparticule{1}{-3}
      \pparticule{7}{3}
   \end{ddiag}
\end{equation}
where circles denote scatterers, horizontal solid lines represent free-space Green functions $G_0$, vertical solid lines
stand for identical scatterers and dashed lines connect different but correlated scatterers. 
The first two terms in \eqref{phase_function_diagrams} correspond to \eqref{phase_function}, and according to the analysis above,
they compensate for a stealth hyperuniform medium provided that $k_0<K/4$. 
The value of the effective scattering mean-free path is obtained by an angular integration (over direction $\bm{u}$) of the phase function~\cite{RYTOV-1989}. 
For ``standard'' short-range order, the first-order correction is obtained by integrating the first two terms, leading to an expression equivalent to that
introduced initially in Ref.~\cite{MARET-1990} to describe multiple scattering in interacting (but non-hyperuniform) suspensions. In the specific case of stealth
hyperuniform disorder, the first two terms do not contribute, and the effective scattering mean-free path is obtained by an angular integration
of the next four diagrams.
For subwavelength scatterers with polarizability $\alpha(\omega)$, this leads to
\begin{equation}\label{ls}
   \frac{1}{\ell}=\frac{8\pi\rho k_0^2}{\ell_B}
      \int_0^{\infty}\Re\left[\alpha(\omega)G_0(r)\right]h(r)
      F(k_0r) r^{d-1}\ud r
\end{equation}
where $h$ is the pair correlation function~\cite{TORQUATO-2003} and $d$ the dimension of space. 
Note that for large scatterers with size comparable with the wavelength (such as Mie spheres), the dependence of 
the scattering amplitude of an individual scatterer on the scattered wavevector $\bm{q}$ would change the analytical expressions.
We can expect in this case the results to change qualitatively.
For scalar waves in 2D, $F(x)=\bessel(x)$ and $G_0(r) = (i/4) \hankel(k_0r)$, with
$\bessel$ ($\hankel$)
the zero-order Bessel (Hankel) function of the first kind. For scalar waves in 3D, $F(x)=2\operatorname{sinc}(x)$
and $G_0(r)=\exp(ik_0r)/(4\pi r)$.  An estimate of the right-hand side can be obtained by computing the
self-energy $\Sigma$.  To first order in $(k_0\ell_B)^{-1}$, one has~\cite{MONTAMBAUX-2007}
\begin{equation}
   \Sigma = 
   \begin{diagc}{2}
      \pparticule{1}{0}
   \end{diagc}
   +
   \begin{diagc}{2}
      \correldeux{1}{7}
      \iidentique{1}{0}{7}{0}
      \pparticule{1}{0}
      \pparticule{7}{0}
   \end{diagc}
\end{equation}
which in Fourier space with $\bm{k}=k_0\bm{u}$ reads
\begin{equation} \label{eq:Sigma_Fourier}
   \Sigma=\rho k_0^2\alpha(\omega)
      \left[
         1+\rho k_0^2\alpha(\omega)
            \int_0^{\infty}G_0(r)h(r)F(k_0r) r^{d-1}\ud r
      \right].
\end{equation}
The integral term in \eqref{eq:Sigma_Fourier} is on the order of $(k_0\ell_B)^{-1}$, which can be used to estimate
the right-hand side in \eqref{ls}, leading to $\ell\sim \ell_B\times k_0\ell_B$.  Transparency is observed provided
that $\ell \gg L$, which leads to the criterion
\begin{equation}
b_B\ll k_0\ell_B . 
\label{final_criterion}
\end{equation}
Under this condition, a stealth hyperuniform material becomes transparent for frequencies satisfying $k_0<K/4$. In particluar, this means that
even for $b_B \gg 1$ (corresponding to the multiple scattering regime for a fully disordered material), transparency
can be reached provided that Eq.~(\ref{final_criterion}) is satisfied. Therefore our analysis establishes a criterion for transparency
of a hyperuniform material even at high density, {\it i.e.} far beyond the single-scattering regime.

% Conclusion
% ==========
\section{Conclusion}

In summary, we have shown that materials made of non-absorbing subwavelength scatterers distributed on a stealth hyperuniform
point pattern can be transparent, even in conditions under which, for the same density of scatterers, an uncorrelated
disordered material would be opaque due to multiple scattering. This occurs under very general conditions, that
we have established theoretically for 2D and 3D materials. In this first study, the numerical examples have been restricted to
2D materials (which by themselves are of practical interest) for the sake of computer time,  but there is no fundamental limitation
for the extension of the numerical simulations to 3D. The transparency property of dense hyperuniform media opens new perspectives 
in the engineering of disordered materials combining  specific photonic properties with a high robustness to fabrication errors. More generally,
the design of dense and transparent hyperuniform materials is in principle achievable for any kind of waves,
and the applications of the concept cover many fields of wave physics.

% Acknowledgments
% ===============
\section*{Acknowledgments}

We thank Kevin Vynck for helpful discussions. 

\section*{Funding}

This work was supported by LABEX WIFI (Laboratory of Excellence within the
French Program ``Investments for the Future'') under references ANR-10-LABX-24 and ANR-10-IDEX-0001-02 PSL* and by the
PEPS program of the French CNRS through project LILAS.

% Bibliography
% ============

\end{document}

\title{Supplemental Material for\\
     ``High-density hyperuniform materials can be transparent''}

\author{O. Leseur}
\affiliation{ESPCI ParisTech, PSL Research University, CNRS, Institut Langevin, 1 rue Jussieu, F-75005, Paris, France}
\author{R. Pierrat}
\affiliation{ESPCI ParisTech, PSL Research University, CNRS, Institut Langevin, 1 rue Jussieu, F-75005, Paris, France}
\author{R. Carminati}\email{remi.carminati@espci.fr}
\affiliation{ESPCI ParisTech, PSL Research University, CNRS, Institut Langevin, 1 rue Jussieu, F-75005, Paris, France}
\date{\today}

\begin{abstract}
This document contains a brief description of the method used to generate hyperuniform point patterns, and a derivation
of the relationship between the averaged structure factor and, respectively, the single-scattering diffuse intensity and the phase function.
\end{abstract}

\maketitle

\section{Method to generate hyperuniform point patterns}
% ======================================================

\begin{figure}[htb]
   \centering
   \psfrag{L}[c]{$L$}
   \psfrag{N}[r]{\rotatebox{90}{\parbox{0.2\linewidth}{$N$ points\\in each cell}}}
   \includegraphics[width=\linewidth]{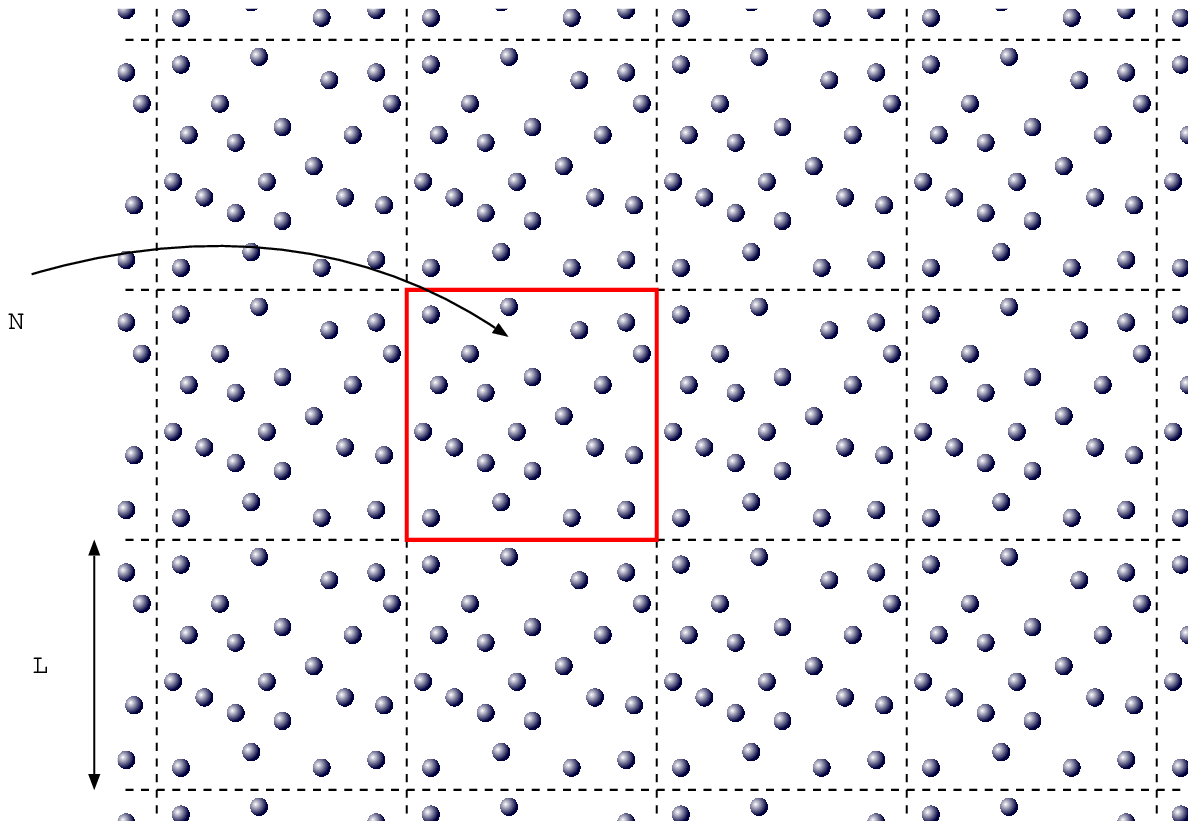}
   \caption{Periodic system considered to generate hyperuniform point patterns.}
   \label{system_perio}
\end{figure}

This section is dedicated to the presentation of the method used to generate 2D hyperuniform point patterns. The algorithm
is adapted from Refs.~\cite{TORQUATO-2004,UCHE-2006,TORQUATO-2008-1}. Numerically we
can only manipulate a finite number of points and to mimic an infinite medium, we divide the space
into identical square cells of size $L$, each containing $N$ points, as shown in \fig{system_perio}. The system is
supposed to be $L$-periodic and we only consider points in one unit cell.

% Figure 1: Periodic system

\begin{figure}[htb]
   \centering
   \psfrag{K}[c]{$K$}
   \psfrag{qx}[c]{$q_x$}
   \psfrag{qy}[c]{$q_y$}
   \psfrag{P}[c]{\parbox{0.2\linewidth}{$(2P+1)^2$\\points}}
   \psfrag{L}[c]{$\displaystyle\frac{2\pi}{L}$}
   \psfrag{O}[c]{$\Omega$}
   \includegraphics[width=0.8\linewidth]{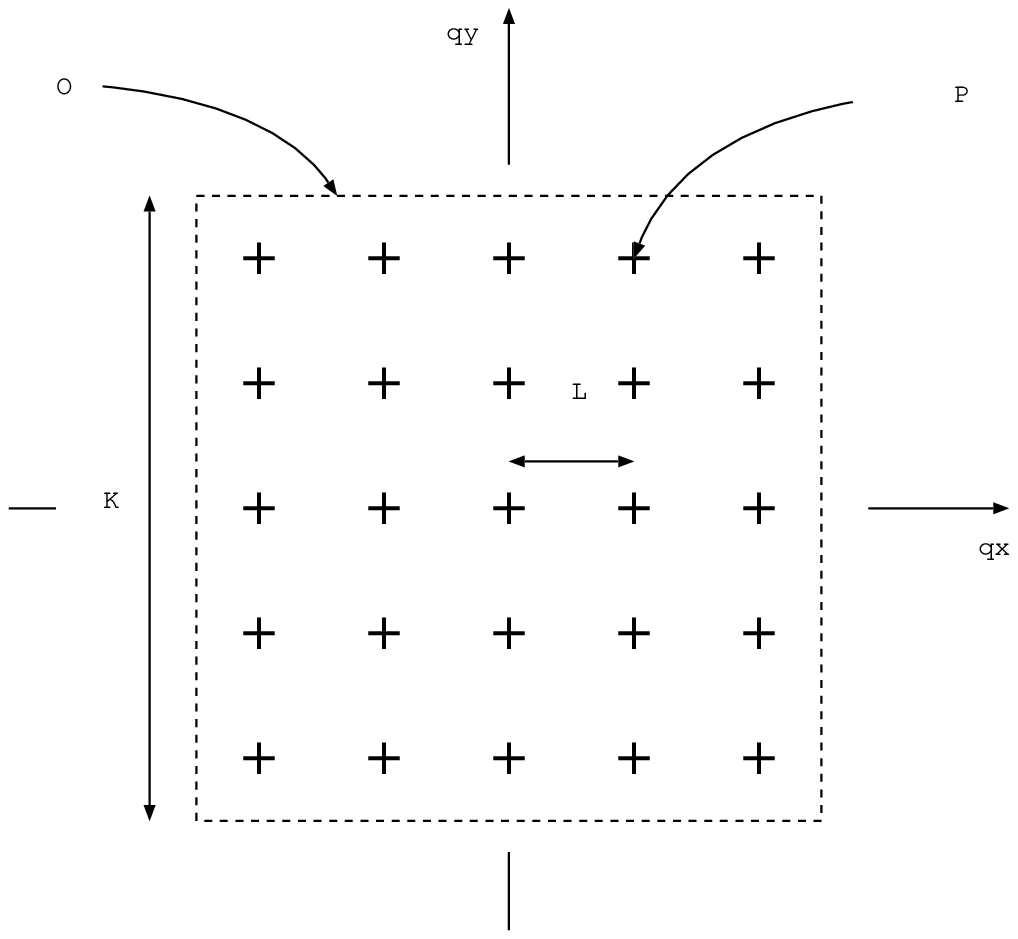}
   \caption{Ensemble $\Omega'$ of points where the structure factor of the elementary cell should vanish to get a
   hyperuniform point pattern.}
   \label{structure_factor}
\end{figure}

The structure factor for the full system is thus given by
\begin{multline}
   \Stot(\bq) = \lim_{R\to\infty}\frac{1}{(2R+1)^2}\frac{1}{N}
\\\times
      \left\lvert\sum_{n_x,n_y=-R}^R\sum_{j=1}^N\exp[i\bq\cdot(n_xL\bm{u}_x+n_yL\bm{u}_y+\br_j)]\right\rvert^2
\end{multline}
where $\bm{u}_x$ and $\bm{u}_y$ are unit vectors along the $x$ and $y$ axes respectively. This leads to the following
expression
\begin{multline}
      \Stot(\bq) = \lim_{R\to\infty}\frac{1}{(2R+1)^2}\left\lvert\frac{\sin[(R+1/2)q_xL]}{\sin(q_xL/2)}\right\rvert^2
\\\times
                                                      \left\lvert\frac{\sin[(R+1/2)q_yL)}{\sin(q_yL/2)}\right\rvert^2
                                    S(\bq)
\end{multline}
where $\bq=(q_x,q_y)$ and
\begin{equation}
   S(\bq)=\frac{1}{N}\left\lvert\sum_{j=1}^N\exp(i\bq\cdot\br_j)\right\rvert^2
\end{equation}
is the structure factor of the unit cell.  This expression implies that $\Stot(\bq)=0$ for all $\bq$ such that $q_x$ and
$q_y$ are not multiple of $2\pi/L$. Thus to get $\Stot(\bq)=0$ for all $\bq\in\Omega$, we only have to require a vanishing of the last
term for any $\bq=(2\pi n/L,2\pi m/L)\in\Omega$, that is
\begin{equation}
   B(\bq)=\left\lvert\sum_{j=1}^N\exp(i\bq\cdot\br_j)\right\rvert^2=0
   \quad\forall\bq\in\Omega'
\end{equation}
where $\Omega'$ is the ensemble corresponding to all $\bq\in\Omega$ such that $\bq=(2\pi n/L,2\pi m/L)$, as shown in
\fig{structure_factor}. $B(\bq)$ is always positive or zero. As a result, to make the structure factor vanishing in
$\Omega'$, $B(\bq)$ must reach its global minimum (\ie zero) for all $\bq\in\Omega'$. The simplest way to implement this
is to define a potential $\phi$ by
\begin{equation}
   \phi(\br_1,\ldots,\br_N) = \sum_{\bq\in\Omega'}B(\bq)
\end{equation}
and to minimize this potential. In the case of a square domain $\Omega$ with size $K$, this leads to the following
expression for $\phi$:
\begin{fmultline}
   \phi(\bm{r}_1,\ldots,\bm{r}_N)=
\\
   \sum_{j=1}^N\sum_{l=1}^N
   \frac{\sin\left[(2P+1)\pi(x_j-x_l)/L\right]}{\sin\left[\pi(x_j-x_l)/L\right]}
\\\times
   \frac{\sin\left[(2P+1)\pi(y_j-y_l)/L\right]}{\sin\left[\pi(y_j-y_l)/L\right]}
\end{fmultline}
where $P=\operatorname{E}[KL/(4\pi)]$.

% Figure 2: Structure factor

\section{Relationship between the average diffuse intensity and the structure factor in the single-scattering regime}
% ===============================================================================================================

In this section, we derive the relationship between the average diffuse intensity and the structure factor in
the single scattering regime. We consider a 3D geometry, but the derivation is essentially the same in 2D 
(and can be easily adapted). 
The average diffuse intensity is given by the difference between the average
intensity and the intensity of the average field (ballistic component) through the relation
\begin{multline}
   I_{\text{diff}}(\theta,\omega)=\bra \lvert E_{\text{sca}}(\theta,\omega)\rvert^2 \ket - \lvert\bra
      E_{\text{sca}}(\theta,\omega)\ket\rvert^2
\\
   \propto \left[\bra\left\lvert\sum_{j=1}^N\exp[i\bm{q}\cdot\bm{r}_j]\right\rvert^2\ket
      -\left\lvert\bra\sum_{j=1}^N\exp[i\bm{q}\cdot\bm{r}_j]\ket\right\rvert^2\right] |E_0|^2
\end{multline}
where $\bm{q}=\bm{k}_s-\bm{k}_i$. The first term corresponds to the average structure factor and is given by
\begin{equation}
   \bra\left\lvert\sum_{j=1}^N\exp[i\bm{q}\cdot\bm{r}_j]\right\rvert^2\ket=N\bra S(\bm{q})\ket.
\end{equation}
The second term is more complex. To have a more explicit expression, we have to consider the probability
$P(\bm{r})=V^{-1}$ of having a scatterer at position $\bm{r}$ which gives
\begin{multline}
   \left\lvert\bra\sum_{j=1}^N\exp[i\bm{q}\cdot\bm{r}_j]\ket\right\rvert^2
      =\left\lvert\sum_{j=1}^N\bra\exp[i\bm{q}\cdot\bm{r}_j]\ket\right\rvert^2
\\\shoveright{
      =\left\lvert\sum_{j=1}^N\int\exp[i\bm{q}\cdot\bm{r}_j]P(\bm{r}_j)\ud^3\bm{r}_j\right\rvert^2
}
\\
      =\frac{N^2}{V^2}\left\lvert\int\exp[i\bm{q}\cdot\bm{r}]\ud^3\bm{r}\right\rvert^2.
\end{multline}
We now define $\Theta$ as
\begin{equation}
   \Theta(\bm{r})=\begin{cases}
      1 & \text{if $\bm{r}\in V$}
   \\
      0 & \text{everywhere else.}
   \end{cases}
\end{equation}
Thus $\Theta$ is a window function, the Fourier transform of which gives the second term in the average diffuse
intensity. Finally, we obtain
\begin{equation}
   \boxed{
      I_{\text{diff}}(\theta,\omega) \propto N\bra S(\bm{q})\ket |E_0|^2
         -\frac{N^2}{V^2}\left\lvert\Theta(\bm{q})\right\rvert^2 |E_0|^2
   }
\end{equation}
which corresponds to Eq.~(4) of the main text.

\section{Relationship between the phase function and the structure factor}
% ====================================================================

In this section, we derive the relationship between the phase function that enters the radiative transfer equation
(RTE)~\cite{CHANDRASEKHAR-1950} and the structure factor in a 3D geometry (the derivation can be easily adapted to 2D). 
We recall that the phase function corresponds to the radiation
pattern of a single scattering event in a macroscopic sense (after averaging over disorder), and
the RTE is a transport equation for the specific intensity (averaged over disorder as well). The RTE can be derived from first principles (Maxwell
equations)~\cite{BARABANENKOV-1968,RYTOV-1989}. More precisely, the starting point of the derivation is 
the Bethe-Salpeter equation~\cite{SHENG-1995,APRESYAN-1996,MONTAMBAUX-2007} which, in statistically homogeneous
medium, reads as
\begin{multline}\label{bethe-salpeter}
   \bra E(\bm{r},\omega)E^*(\bm{\uprho},\omega)\ket=
      \bra E(\bm{r},\omega)\ket\bra E^*(\bm{\uprho},\omega)\ket
\\
      +\int \bra G(\bm{r}-\bm{r}',\omega)\ket\bra G^*(\bm{\uprho}-\bm{\uprho}',\omega)\ket
      \Gamma(\bm{r}',\bm{r}'',\bm{\uprho}',\bm{\uprho}'',\omega)
\\\times
      \bra E(\bm{r}'',\omega)E^*(\bm{\uprho}'',\omega)\ket
      \ud^3\bm{r}'\ud^3\bm{r}''\ud^3\bm{\uprho}'\ud^3\bm{\uprho}''
\end{multline}
where $\bra\ldots\ket$ denotes the ensemble average over the configurations of disorder, $\bra
E(\bm{r},\omega)E^*(\bm{\uprho},\omega)\ket$ being the spatial correlation of the field and $\bra
G(\bm{r}-\bm{r}',\omega)\ket$ the average Green function. $\Gamma$ is called the intensity vertex operator. It is a very complex
object that contains all multiple scattering events that cannot be factorized when averaging over disorder. In particular,
$\Gamma$ contains the information on the spatial correlations of the disrodered medium.
From \eqref{bethe-salpeter}, and considering a dilute system with $k_0\ell_B\gg 1$,
one can derive the RTE, and in particular the expression of the phase function that reads
\begin{equation}
   p(\bm{u},\bm{u}')=A\widetilde{\Gamma}(k_r\bm{u},k_r\bm{u}',k_r\bm{u},k_r\bm{u}',\omega)
\end{equation}
where the constant factor $A$ is such that $\int p(\bm{u},\bm{u}')\ud\bm{u}=4\pi$.
$k_r=\sqrt{\Re\left[k_{\text{eff}}^2\right]}$ where $k_{\text{eff}}=k_0+i/(2\ell)$ is the effective wave vector (\ie the
wave vector appearing in the wave equation for the average field). In a dilute system, we have $k_r\sim k_0$.
$\widetilde{\Gamma}$ is the reduced vertex operator in Fourier space, given by
\begin{multline}
   \Gamma(\bm{k},\bm{k}',\bm{\upkappa},\bm{\upkappa}',\omega)=(2\pi)^3\delta(\bm{k}-\bm{k}'-\bm{\upkappa}+\bm{\upkappa}')
\\\times
      \widetilde{\Gamma}(\bm{k},\bm{k}',\bm{\upkappa},\bm{\upkappa}',\omega).
\end{multline}
In a dilute system, it is convenient to use a Taylor expansion of $\Gamma$ with respect to the small parameter $[k_0\ell_B]^{-1}$.
In the case of a correlated medium (such as a hyperuniform material), the first two terms are needed, and the expansion is
\begin{equation}
   \Gamma=
   \begin{ddiag}{2}
      \iidentique{1}{3}{1}{-3}
      \pparticule{1}{-3}
      \pparticule{1}{3}
   \end{ddiag}
   +
   \begin{ddiag}{2}
      \ccorreldeuxc{1}{3}{1}{-3}
      \pparticule{1}{-3}
      \pparticule{1}{3}
   \end{ddiag}
\end{equation}
where the first term corresponds to a scattering process for the field and its complex conjugate involving the same scatterer (solid line
between both circles representing the scatterers, the upper row being related to the field and the lower row to its
conjugate). The second term describes a scattering process involving two different but correlated scatterers
(dashed line). Analytically, for subwavelength satterers with polarizability $\alpha(\omega)$, we obtain
\begin{multline}
   \Gamma(\bm{r}',\bm{r}'',\bm{\uprho}',\bm{\uprho}'',\omega)
      =\frac{N}{V}k_0^4\lvert\alpha(\omega)\rvert^2\delta(\bm{r}''-\bm{r}')\delta(\bm{\uprho}''-\bm{\uprho}')
\\\times
      \left[
         \delta(\bm{\uprho}'-\bm{r}')
         +\frac{N-1}{V}h(\bm{\uprho}'-\bm{r}')
      \right]
\end{multline}
where $h$ is the pair correlation function~\cite{TORQUATO-2003}. This leads to the following expression for the phase
function:
\begin{equation}
   p(\bm{u},\bm{u}')=\frac{AN}{V}k_0^4\lvert\alpha(\omega)\rvert^2\left\{1+\frac{N-1}{V}h[k_0(\bm{u}-\bm{u}')]\right\}.
\end{equation}
To get the relationship with the structure factor, we now explicitely write $\bra S\ket$ in terms of the pair correlation function. We have
\begin{align}\nonumber
   \bra S(\bm{q})\ket & =\frac{1}{N}\bra\left|\sum_{j=1}^N\exp\left[i\bm{q}\cdot\bm{r}_j\right]\right|^2\ket
\\\nonumber
   & = 1+\frac{1}{N}\bra\sum_{j=1}^N\sum_{\substack{k=1\\k\ne j}}^N
         \exp\left[i\bm{q}\cdot(\bm{r}_j-\bm{r}_k)\right]\ket
\\\nonumber
   & = 1+\frac{1}{N}\int_{V^2}\sum_{j=1}^N\sum_{\substack{k=1\\k\ne j}}^N
         \exp\left[i\bm{q}\cdot(\bm{r}_j-\bm{r}_k)\right]
\\\nonumber & \hphantom{= 1+}\times
         P(\bm{r}_j)P(\bm{r}_k)[1+h(\bm{r}_j,\bm{r}_k)]\ud^3\bm{r}_j\ud^3\bm{r}_k
\\\nonumber
   & = 1+\frac{N-1}{V^2}\int_{V^2}\exp\left[i\bm{q}\cdot(\bm{r}_1-\bm{r}_2)\right]
\\\nonumber & \hphantom{= 1+}\times
      [1+h(\bm{r}_1-\bm{r}_2)]\ud^3\bm{r}_1\ud^3\bm{r}_2
\\
   & = 1+\frac{N-1}{V}h(\bm{q})+\frac{N-1}{V^2}\lvert\Theta(\bm{q})\rvert^2.
\end{align}
From Eqs. (17) and (18), we finally obtain Eq.~(5) of the main text:
\begin{fequation}
   p(\bm{u},\bm{u}')\propto\bra S[k_0(\bm{u}-\bm{u}')]\ket
      -\frac{N-1}{V^2}\left\lvert\Theta[k_0(\bm{u}-\bm{u}')]\right\rvert^2.
\end{fequation}
The appearence of the correction term in the right-hand side results from the definition of the phase function, that does not involve the same diagrams
as the structure factor. In the phase function, or equivalently in the intensity vertex, we only take into account
irreductible diagrams (\ie diagrams that can not be factorized after the average process). In the structure factor, we
have an additionnal diagram corresponding to two different scatterers that are not correlated, so that the expression of the averaged structure factor reads
reads
\begin{equation}
   \bra S\ket=
   \begin{ddiag}{2}
      \iidentique{1}{3}{1}{-3}
      \pparticule{1}{-3}
      \pparticule{1}{3}
   \end{ddiag}
   +
   \begin{ddiag}{2}
      \ccorreldeuxc{1}{3}{1}{-3}
      \pparticule{1}{-3}
      \pparticule{1}{3}
   \end{ddiag}
   +
   \begin{ddiag}{2}
      \pparticule{1}{-3}
      \pparticule{1}{3}
   \end{ddiag}.
\end{equation}

% Bibliography
% ============